\documentclass{article}

\usepackage{microtype}
\usepackage{appendix}
\usepackage{graphicx}
\usepackage{booktabs} % for professional tables
\usepackage{MnSymbol,wasysym}
\usepackage{pifont}% http://ctan.org/pkg/pifont
\newcommand{\cmark}{\ding{51}}%
\newcommand{\xmark}{\ding{55}}%
\usepackage{etoolbox}
\usepackage{siunitx}
\usepackage{enumitem}
\usepackage{caption}
\usepackage[small,compact]{titlesec}
\usepackage{amsmath}
\usepackage{algorithmic}
\usepackage{textcomp}
\usepackage[dvipsnames]{xcolor}
\usepackage{lipsum}
\usepackage[first=1,last=100]{lcg}
\usepackage{blindtext}
\usepackage{adjustbox}
\usepackage{multicol}

\usepackage{enumitem}
\setlist{nolistsep}

\renewcommand\fbox{\fcolorbox{gray}{white}}

\newcommand{\camera}[1]{\textcolor{black}{#1}}

\usepackage{multirow}

\usepackage{hyperref}

\usepackage{url}

\usepackage[accepted]{mlsys2023}
\begin{document}

\newcommand\nnfootnote[1]{%
  \begin{NoHyper}
  \renewcommand\thefootnote{}\footnote{#1}%
  \addtocounter{footnote}{-1}%
  \end{NoHyper}
}
\twocolumn[
\mlsystitle{Edge Impulse: An MLOps Platform for Tiny Machine Learning}

\raggedbottom

% unless you've provided the [accepted] option to the mlsys2023


\mlsyssetsymbol{equal}{*}

\begin{mlsysauthorlist}

% \mlsysauthor{name}{org}

\mlsysauthor{Shawn Hymel}{equal}
\mlsysauthor{Colby Banbury}{equal}
\mlsysauthor{Daniel Situnayake}{}
\mlsysauthor{Alex Elium}{}
\mlsysauthor{Carl Ward}{}
\mlsysauthor{Mat Kelcey}{}
\mlsysauthor{Mathijs Baaijens}{}
\mlsysauthor{Mateusz Majchrzycki}{}
\mlsysauthor{Jenny Plunkett}{}
\mlsysauthor{David Tischler}{}
\mlsysauthor{Alessandro Grande}{}
\mlsysauthor{Louis Moreau}{}
\mlsysauthor{Dmitry Maslov}{}
\mlsysauthor{Artie Beavis}{}
\mlsysauthor{Jan Jongboom}{}
\mlsysauthor{Vijay Janapa Reddi}{}

\end{mlsysauthorlist}
% \mlsysaffiliation{}{}
% \mlsysaffiliation{harvard}{Harvard University}
% \mlsysaffiliation{edgeimpulse}{Edge Impulse}

% \mlsyscorrespondingauthor{Colby Banbury}{cbanbury@g.harvard.edu}

\mlsyskeywords{Machine Learning, MLSys, embedded ML, edge ML, TinyML}

\vspace*{1em}

\begin{abstract}

Edge Impulse is a cloud-based machine learning operations (MLOps) platform for developing embedded and edge ML (TinyML) systems that can be deployed to a wide range of hardware targets. Current TinyML workflows are plagued by fragmented software stacks and heterogeneous deployment hardware, making ML model optimizations difficult and unportable. We present Edge Impulse, a practical MLOps platform for developing TinyML systems at scale. Edge Impulse addresses these challenges and streamlines the TinyML design cycle by supporting various software and hardware optimizations to create an extensible and portable software stack for a multitude of embedded systems. As of Oct. 2022, Edge Impulse hosts 118,185 projects from 50,953 developers.

\end{abstract}

]
% \printAffiliationsAndNotice{\mlsysEqualContribution}
\nnfootnote{
*Equal Contribution\\
Corresponding author:
\textless{}cbanbury@g.harvard.edu\textgreater{}. \\ 
Colby Banbury and Vijay Janapa Reddi are with the John A. Paulson School of Engineering and Applied Sciences, Harvard University. All others are with Edge Impulse.\\ 
Edge Impulse website: \url{http://www.edgeimpulse.com/}\\
\textit{Proceedings of the
		$\mathit{6}^{th}$ MLSys Conference},
	Miami Beach, FL, USA, 2023.
  Copyright 2023 by the author(s).}


%\printAffiliationsAndNotice{} % otherwise use the standard text.

\vspace{-20pt}
\section{Introduction}
\label{sec:introduction}

Machine learning (ML) has become an increasingly important tool in embedded systems for solving difficult problems, enhancing existing Internet of Things (IoT) infrastructure, and offering unique ways to save power and bandwidth in sensor networks. ML inference on \textit{TinyML} systems has facilitated the development of technologies in low-power devices such as wakeword detection~\cite{gruenstein2017cascade}, predictive maintenance~\cite{susto2015mlforpm}, anomaly detection~\cite{koizumi2019toyadmos}, visual object detection~\cite{chowdhery2019visualwakewords}, and human activity recognition~\cite{chavarriaga2013benchmark}.
According to ABI Resarch, a global technology intelligence firm, the ``installed base of devices with edge AI chipset will exceed 5 billion by 2025.'' Additionally, the embedded ML market is expected to reach US\$44.3 billion by 2027~\cite{abiresearch2021edgeai}.

Despite the promising advances, the embedded ML development ecosystem has lagged behind the demand for applications. The embedded ML development workflow often requires specific expertise. For instance, embedded ML developers often have to learn a new set of tools for training new models and porting them to an embedded framework written in C or C++, while managing conflicting library dependencies. Additionally, hardware vendor specific frameworks often lock a developer into a particular ecosystem, which limits the flexibility and scalability of an application. 
% As IoT infrastructure grows, the need for scalable embedded ML operations (MLOps) will also grow.
% , which involves multiple interconnected processes.

% TensorFlow Lite Micro (TFLM) \cite{david2021tensorflow} is an open-source framework for providing ML inference of deep-learning models on embedded systems. Once model training has been performed inside TensorFlow, the model can be translated into a buffer that can be read by the TFLM framework using C++. At the time of writing, TFLM supports only a subset of TensorFlow operations, including, among others, dense neural network (NN) layers and convolutional layers. However, this limited subset of operations allows embedded developers to perform model training on larger, more powerful computers and inference on a wide range of devices, including microcontrollers. 

\begin{figure*}
    \centering
    \includegraphics[width=\textwidth]{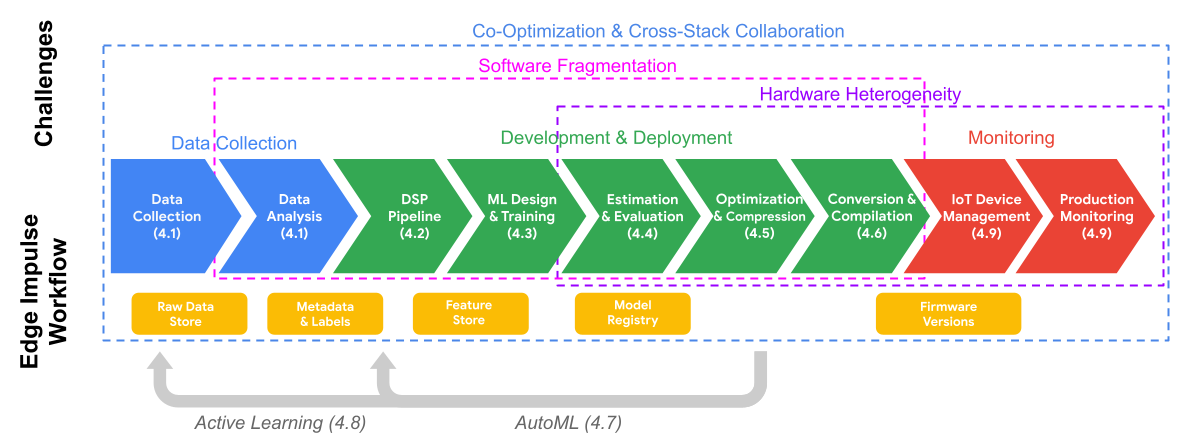}
    \caption{The challenges associated with the ML Workflow and features of Edge Impulse that solve those challenges.}
    \label{fig:ml-workflow}
\end{figure*}

While popular frameworks such as
% Google's 
TensorFlow Lite for Microcontrollers (TFLM)~\cite{david2021tensorflow} help address optimization and compression of neural networks for embedded devices,
% academic and industry
adoption has been slow due to challenges that are unique to the {\textit{embedded} machine learning ecosystem}. Broadly, these include the following:
\begin{enumerate}
\itemsep0em 

    \item \textit{Data collection challenge.} There is no large-scale, curated, public sensor data set for the embedded ecosystem. Currently, it is difficult to efficiently collect, and analyze such datasets from a rich variety of sensors.
    Additionally, data cleaning and labeling are essential to ML development but are expensive, labor intensive processes without tooling or automation.
    
    \item \textit{Data preprocessing challenge.} 
    Digital signal processing (DSP) is a critical stage of the ML stack and has strong interactions with the ML model, that are sometimes hard to quantify.
    Yet there are a lack of automated machine learning tools for the embedded ecosystem that include the DSP component, which hinders the development of efficient preprocessing methods for these systems by non-domain experts.
    
    \item \textit{Development challenge.}
    % In practice, the TensorFlow and TFLM ecosystem make model training and development straightforward. However, 
    Matching TensorFlow and TFLM dependency versions across training and deployment infrastructure is challenging. To ensure functional reliability, it is often necessary for developers to have extensive knowledge in multiple domains, such as Python, machine learning, TensorFlow, C/C++, and embedded systems. 
    
    \item \textit{Deployment challenge.} Scaling ML deployment is hampered by the rich heterogeneity of embedded architectures and development frameworks, which restricts code portability, particularly as a result of architecture-specific optimization strategies. In addition, the absence of automated machine learning (AutoML) tools to assist non-domain experts in developing model architectures for embedded systems restricts accessibility.
    
    \item \textit{Monitoring challenge.} In contrast to traditional cloud-based machine learning systems that rely on mature software and hardware stacks, there is no unified MLOps framework for programmatically updating datasets, training models, and deploying them to embedded devices. Additionally, few benchmarking tools exist to quantify model performance on diverse embedded architectures that are highly heterogeneous.
\end{enumerate}

Edge Impulse, an online platform designed to simplify the process of collecting data, train deep learning models, and deploy them to embedded and edge computing devices, allows us to address these aforementioned issues. Edge Impulse targets customers in the business sector who want to develop edge machine learning (ML) solutions for a variety of problems. However, the Edge Impulse platform also facilitates a research- and classroom-friendly environment. 

Figure~\ref{fig:ml-workflow} illustrates the end-to-end ML workflow of Edge Impulse.
% It has a rich array of features.
Edge Impulse simplifies the process of data collection and curation for users and streamlines the training and evaluation of models. Users can interact with the training and deployment process via a combination of a web-based graphical user interface (GUI) and an API. Edge Impulse also provides an extensible and portable C/C++ library that encapsulates the preprocessing code and trained model to make inferencing simple across a wide range of target devices as well as a number of target-specific optimizations to reduce inference time and model memory consumption.

Edge Impulse offers several key technical contributions that are unique. The first contribution is a data collection system that helps users collect and store training and test data alongside their model and deployment code. Rather than relying on prebuilt datasets or requiring users to construct their own data gathering technology, Edge Impulse offers a variety of methods to gather data in real-world environments. The second contribution is pairing preprocessing feature extraction with deep learning, which allows users to explore a range of possible solutions to their individual problem or task. The Edge Optimized Neural (EON) Tuner assists in this task by automatically exploring a user-defined search space of both preprocessors and ML models. The third contribution is an extensible and portable inferencing library that can be deployed across a wide range of edge and embedded systems. The EON Compiler removes the overhead required by the TFLM interpreter, thereby reducing usage of the limited RAM and flash space.

In this paper, we outline the challenges of developing and deploying ML models to embedded devices from an industry practice perspective. Next, we describe the architecture and use cases for a platform designed specifically to address these obstacles. We provide several examples to illustrate how this platform has been utilized successfully in industry, academia, and research institutions to develop novel machine learning-based solutions. Finally, we provide an evaluation of the performance and portability of the platform-generated inference code.
\section{Embedded Ecosystem Challenges}
\label{sec:challenges}

% In TinyML, applications are frequently defined by their constraints. The computational capabilities, memory capacity, and energy availability of embedded devices are severly constrained, and the complexity and nacency of the deployment stack make it difficult to develop TinyML applications. 

In Section~\ref{sec:introduction}, we highlight the challenges faced by an embedded ML developer. In this section, we highlight the challenges posed by the Embedded ecosystem which make platform and framework development difficult.
% associated with developing applications for severely constrained hardware and managing a rapidly evolving embedded ecosystem.

\subsection{Device Resource Constraints}
\label{sec:constraints}

TinyML systems often have very limited computational capabilities, due to their small size, cost, and energy budget. Microcontrollers, which are the most common and general purpose processors in the TinyML space, often have much lower clock speeds and fewer architectural features~\cite{banbury2021micronets} than their mobile or server-class counterparts. This becomes a challenge when trying to keep pace with the flow of data from a sensor or hit latency constraints.

% More recently, some TinyML systems have integrated accelerators to speed up computation, such as digital signal processors~cite(), neural engines~cite(), or custom function units~cite(). While these application specific circuits alleviate some of the compute constraints, they can add complexity to an already complicated deployment stack,

ML workloads often require gigabytes of working memory and storage to store activations and model weights, but TinyML systems are often equipped with only a few hundred kilobytes of SRAM and a few megabytes of eFlash. This enforces a strict constrain on the models. TinyML systems often have very flat memory hierarchies, due to small or non-existent caches and often no off-chip memory~\cite{banbury2021micronets}. This means the typical data access patterns that neural networks have been designed around no longer apply, which has forced the design of new model architectures~\cite{banbury2021micronets,lin2020mcunet}.

% \subsubsection{Storage}
% Microcontrollers and other TinyML systems often come equipped with a few gigabytes of embedded flash storage~\cite{}, which needs to store all application code in addition to the model’s parameters. This imposes a hard constraint on the size and depth of an embedded model. Additionally, for microcontrollers, the flash memory is often on the same chip as the processor, which means storage access times are often lower, relative to the compute, than mobile class processors ~\cite{micronets}. This means that fewer of the model’s weights needs to be pre-loaded into working memory, reducing the pressure on the memory capacity. 
% \subsubsection{Energy}
Finally, many TinyML applications operate on battery power, and the battery life of the system directly impacts the usefulness of the application. Due to the small size and cost of TinyML systems, these batteries are often small and low capacity (e.g. a coin cell). 
% Therefore reducing energy consumption is a priority for TinyML optimization.
Due to the limited energy budget, any wireless transmission can quickly deplete the battery~\cite{siekkinen2012low}. Since data is often only transmitted once a specific prediction is made (e.g. ``OK Google'', ``Alexa'', ``Hey Siri'', etc.), false positives contribute to battery drain with no benefit. Therefore, the accuracy of a model can directly impact the energy consumption of the system. 
% Furthermore, if the model latency is well below the application requirement, the device can enter sleep mode between inferences to conserve energy.
These device constraints force TinyML application developers to leverage every compression and optimization technique at their disposal, which, as described in the next sections, poses it's own  challenges.
% \subsection{Development and Deployment Challenges}

\subsection{Hardware Heterogeneity}
\label{hardware-het}

Despite resource limitations being fairly constant across TinyML hardware, the embedded computing systems themselves are quite diverse. TinyML devices range from microcontrollers~\cite{eggimann2019risc} and digital signal processors~\cite{gruenstein2017cascade}, to application specific accelerators~\cite{prakash2022cfu, ethos-u65} and neurmorphic processors~\cite{qiao2015reconfigurable}. 
The STM32 32-bit Arm Cortex MCU family alone, for example, includes 17 series of microcontrollers.
% H7, F7, F4, F3, F2, F1, F0, G4, G0, L5, L4, L4+ L1, L0, U5, WL, and WB.
Each STM32 microcontroller series is based on an ARM processor core that is either Cortex-M7, Cortex-M4, Cortex-M33, Cortex-M3, Cortex-M0+, or Cortex-M0~\cite{STM32Arm35:online}. Their capabilities can also vary at the instruction set architecture level. The same is true of other vendors.
Each hardware platform supports different deployment processes, model types, numerical formats, and memory access patterns, which often makes TinyML applications difficult to port across devices. This complexity is exacerbated when a creating an application at scale, which must be deployed to a wide variety of devices, each with their own libraries and deployment method.

\subsection{Software Fragmentation}
\label{software-frag}

Due to TinyML's infancy, the software stack has not yet reached a state of stability concerning particular formats and best practices. Occasionally, TinyML applications are deployed with a full operating system (OS) like Linux, a real-time OS like Zephyr~\cite{ZephyrPr75:online}, an inference framework like TFLM~\cite{david2021tensorflow}, or even a bare-metal implementation as a C++ library with no external dependencies.
% which can be compiled with any modern C++ compiler.
This diversity restricts the interoperability of new optimizations and tools
% at the model level. 
A standard TinyML training pipeline incorporates tools and techniques from multiple sources, resulting in a tangled web of software versions and ports that can hinder collaboration, portability, robustness, and reproducibility.

% \textbf{concrete examples of optimization tools}

\subsection{Co-Optimization and Cross-Stack Collaboration}
\label{dev-deploy-challenges}

Each software and hardware optimization depends on the other layers of the development and deployment stack to be effective. This necessitates a complex optimization problem with many interconnected knobs to tune for optimal performance, as TinyML applications have stringent constraints. 

Even without the addition of model hyper-parameters and optimizations such as quantization, applying consistent pre-processing across projects is a complicated jumble of hyperparameters that often requires deep, domain-specific insights into a signal. Due to this inherent complexity, TinyML development is a time consuming process that requires a wide range of specific technical expertise that is often not readily available in the industry. In addition, the cross-product of options and versions at each layer complicates collaboration and the reproducibility of ML applications. 

Additionally, data consistency poses it's own challenges, especially when using an internal or collected dataset. Significant operational challenges are posed by maintaining train, validation, and test splits, adding or removing individual samples, and preserving metadata. To facilitate large-scale collaborative projects and aid in the resolution of the ML reproducibility crisis~\cite{hutson2018artificial}, one must version control the data, preprocessing, model, and deployment code while tracking a complex web of external dependencies.
\section{Overview and Design Objectives}
\label{sec:edgeimpulse}

Edge Impulse is a combination of software-as-a-service, developer tooling, embedded software, and documentation to help embedded development teams create software that makes use of embedded machine learning at scale. At the time of writing, Edge Impulse is being used by 50,953 developers in 118,185 projects, of which, 3,219 have been made public; it is in use at over 5500 enterprise organizations, excluding universities and other educational institutions.

\begin{figure*}
    \centering
    \fbox{\includegraphics[width=.99\textwidth]{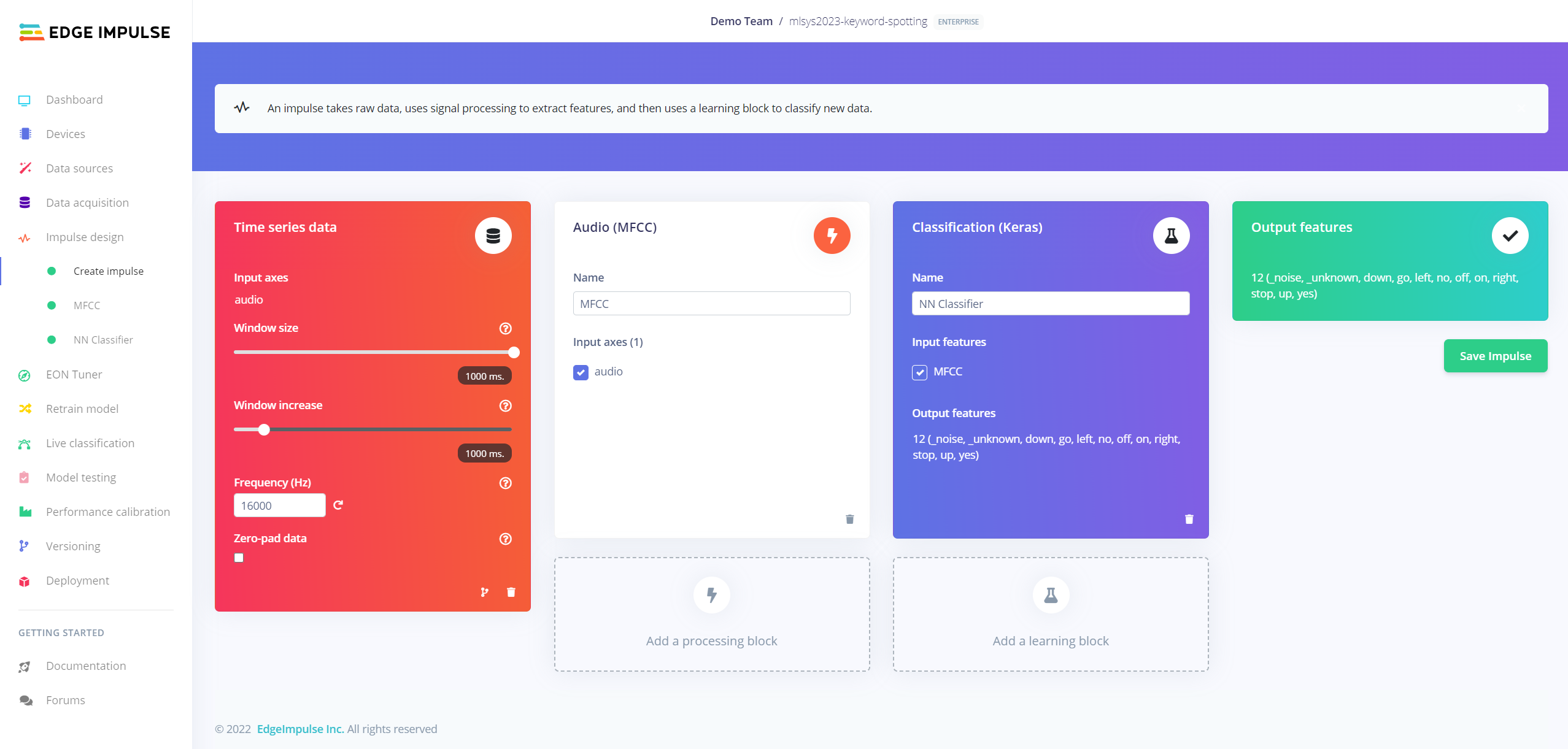}}
    \caption{Screenshot showing the user's view inside an Edge Impulse project where the  \textit{blocks} are connected depicting the dataflow.}
    \label{fig:ei-screenshot}
\end{figure*}

Edge Impulse is designed and engineered according to the following seven guiding principles, based on the developer challenges as well as the embedded ecosystem challenges that we described in Section~\ref{sec:introduction} and Section~\ref{sec:challenges}, respectively.

\textbf{1. Accessible.} Edge Impulse's primary objective is to make embedded machine learning simpler and more accessible while focusing on producing Tiny ML solutions for resource-constrained devices (Section~\ref{sec:constraints}). 
This effectively broadens the pool of potential embedded ML developers by helping the embedded engineers with ML and the ML engineers with embedded systems (Section~\ref{dev-deploy-challenges}).
% This objective is primarily for the benefit of embedded software developers and machine learning engineers. With Edge Impulse, embedded software developers can execute the technical steps of the machine learning workflow without prior knowledge or training in machine learning or data engineering toolchains. Similarly, Edge Impulse allows ML engineers to develop models for embedded targets without prior knowledge or training in embedded toolchains.

\textbf{2. End-to-end.} Edge Impulse provides users the ability to easily experiment with the end-to-end ML workflow holistically (Figure~\ref{fig:ml-workflow}).
% including dataset construction, feature engineering, machine learning and deployment optimizations.
Using Edge Impulse, one (or a team) could collect a dataset, train a efficient, optimized model,
% that is compatible with a specific embedded target
evaluate its performance,
% on realistic data
and deploy embedded firmware.
% Edge Impulse is designed to function both as a stand-alone end-to-end framework and in conjunction with other tools via multiple integration points. Such extensibility enables development across languages and platforms to help alleviate the software fragmentation issues (Section~\ref{software-frag}).

\textbf{3. Data-centric.} Edge Impulse prioritizes a data-centric approach because data collection and analysis has been historically slowed down in the ML pipeline. Given the scarcity of sensor datasets in the embedded ecosystem (Challenge \#1, Section~\ref{sec:introduction}), Edge Impulse enables users to ingest data from various sources. Therefore, Edge Impulse encourages a data-centric approach to ML development, rather than (over) emphasizing a model-centric approach.

\textbf{4. Iterative.} Since cross-stack optimization is critical (Section~\ref{dev-deploy-challenges}), Edge Impulse promotes short developer feedback loops that allows developers to quickly experiment and iterate over different design space optimizations. To this end, Edge Impulse strives to provide a rich set of AutoML tools. Short design cycles and AutoML tools removes some of the burden of expertise (Section~\ref{dev-deploy-challenges}).

\textbf{5. Easy integration and extensible.} Edge Impulse prioritizes integration and extensibility to  address the challenge of software fragmentation (Section~\ref{software-frag}) and cross-stack collaboration (Section~\ref{dev-deploy-challenges}). Experts should be able to connect the technology with their preferred downstream stacks, ideally using open standards where possible, and deploy to a wide variety of embedded and edge platforms (Section~\ref{hardware-het}).

\textbf{6. Team Oriented.} To scale well (Challenge \#5, Section~\ref{sec:introduction}), Edge Impulse facilitates the teamwork and communication required for many embedded machine learning projects by supporting multiple users on projects, versioning of projects, and sharing of projects. In addition, it is well-documented with accessible content that serves every user type.

\textbf{7. Community supported.} Finally, the technology should promote a strong commitment to the community and exist within an ecosystem of community users, tools, and content.

When a user creates a project, they are guided through the process of gathering data, analyzing that data, creating a DSP preprocessing block, training a machine learning model, evaluating that model, and ultimately deploying it to a hardware platform of their choice. These steps are shown in the ML workflow in Figure~\ref{fig:ml-workflow}. Figure~\ref{fig:ei-screenshot} shows a user's view inside an Edge Impulse Studio project with the ML workflow steps shown on the left side of the page. Projects in Edge Impulse are divided into a series of \textit{blocks} that represent the dataflow. For this keyword spotting example, data arrives (i.e. from a microphone) in the left block labeled ``Time series data'' is preprocessed into Mel-frequency cepstral coefficients (MFCCs) in the middle block, and then sent to a NN for inference in the block labeled ``Classification (Keras).'' In the rest of the project design, users can modify block parameters to adjust functionality or create their own blocks to transform the data.
s

With the design objectives established, a few things are beyond the scope of Edge Impulse. Edge Impulse is not intended to eliminate the need for a design process informed by domain expertise, stakeholder consultation, and machine learning workflow insight. For instance, a team of engineers utilizing Edge Impulse must still assess the suitability of machine learning as a solution to the problem they are attempting to solve. The team must have the necessary domain expertise to comprehend the problem and develop a responsible solution. They must still comprehend the general nature of the machine learning (ML) workflow, including iterative development and adopt appropriate evaluation metrics.

\section{Implementation}

In this section, we describe all the different aspects of the end-to-end flow that Edge Impulse supports, as illustrated in Figure~\ref{fig:ml-workflow}. For each stage, we present the rationale for the stage and describe its implementation specifics. 

\subsection{Data Collection and Analysis}
 
Since every ML project begins with data that is often hard to gather easily, Edge Impulse provides a number of features designed to help users collect data, manage their dataset, and perform feature extraction through digital signal processing (DSP). Edge Impulse projects can accept data stored in a several file formats: CSV, CBOR, JSON, WAV, JPG, or PNG. The platform also offers several methods to help users gather data for their project, including command line interface (CLI) tools that interface with device firmware to ingest data in real time and web-based API to upload data directly or from an existing cloud-based store (e.g. AWS S3 bucket). A GUI allows users to visualize training and test set split as well as class allocation grouped into buckets. Users can also examine raw data in each sample through time series plots or images, depending on the data type.

\subsection{DSP Pipeline}

Many of the embedded machine learning applications or use cases rely on sensor data preprocessing. Edge Impulse offers the ability to perform various preprocessing of raw signal data automatically prior to use in training or inference. This preprocessing step is known as \textit{digital signal processing (DSP)} in the Edge Impulse workflow. By preprocessing raw data, model size can often be reduced, and preprocessing can incorporate more efficient algorithms than can be realized with typical neural net architectures. For example, an FFT is an $O(n \cdot log(n))$ algorithm for extracting frequency information, whereas using 1D convolutional layers to accomplish the same thing would require $O(n^{2})$ operations.

Edge Impulse simplifies the preprocessing workflow by providing a rich array of continuum blocks that trade off model size and complexity~\cite{edgeimpu21:online}, a visual explorer for tuning DSP block hyperparameters (frame length, stride, window size, number of coefficients, etc.), and estimates of memory and latency requirements for a given choice of hyperparameters. 

Edge Impulse offers sensible defaults for a variety of tasks to ensure minimal knowledge is required by users, though domain experts can choose preprocessing steps and hyperparameters that reduce the amount of training noise. 
\camera{Additionally, users can automatically select these hyperparameters via the DSP autotune feature, or optimize them via the Eon Tuner (Sec. \ref{auto-ml})}
% Such preprocessing is equivalent to feature engineering that assists edge ML practitioners in choosing the best data pipeline for their model within their latency and memory constraints.

\subsection{ML Design and Training}

Traditionally, a user would need to write code to train a neural network on their data. But to make machine learning more accessible to everyone, Edge Impulse employs a visual editor that allows a user to train on their data without entering any code. There are preset neural network architectures that are suggested based on the type of data coming into the machine learning block. However, the layers or the network can be customized by the user. Advanced users can download the containerized code for the block to train locally or use expert mode to customize further using the Keras framework~\cite{chollet2015keras} and Python.

%, which is a subset of machine learning named after the increased number of layers of artificial neural networks that can now be used as universal function approximators

Arbitrary combinations of building blocks (as shown in the Figure~\ref{fig:ei-screenshot} screenshot) allows for rich flexibility in model architecture, but it is important that the model is trainable. Edge Impulse provides a number of subtle, but important, optimisation pieces to ensure stable training including, but not limited to, learning rate finding, classifier bias initialisation, best model checkpoint restoration. Such building blocks and optimizations make the training process accessible to machine learning novices while the extinsibility of the expert mode allow domain experts to develop more complex ML models on Edge Impulse.

%One benefit of neural networks is that the weights learned by the network during training are transferable to a network for a completely different task in a process called transfer learning \cite{pan2009survey}. This technique can significantly decrease the amount of training data and time required by a user, thus simplifying the data collection and model training steps. 

Edge Impulse provides a transfer learning~\cite{pan2009survey} block for audio keyword detection.
% that users can apply to audio data the same transfer learning technique used for image classification, based on the MobileNetV1~\cite{howard2017mobilenets} and V2 architectures~\cite{sandler2018mobilenetv2}. 
This allows the users to quickly develop a robust keyword spotting application, even when working with a relatively small dataset.

Edge Impulse maintains partnerships with silicon vendors who have developed specific neural network accelerator hardware, such as the Syntiant NDP101. In addition to generating models for general purpose processors, Edge Impulse supports a variety of architecture-specific devices and optimizations, such as CMSIS-NN~\cite{lai2018cmsisnn} to maximize the performance and minimize the memory footprint of neural networks on Cortex-M processor cores, thus alleviating many issues found with hardware heterogeneity.

Edge Impulse also supports several unsupervised learning algorithms to tackle anomaly detection problems. At the moment, Edge Impulse uses K-means clustering and will support Gaussian mixture models (GMM) in the near future.

\subsection{Estimation and Evaluation}
\label{sec:estimation_evaluation}

Since embedded systems are resource-constrained, developers can benefit from having estimations of model inference latency time, RAM usage, and flash memory usage during the early-stage design space exploration. Edge Impulse uses Renode \cite{holenko2015renode} and device-specific benchmarking to produce estimates of preprocessing and model inference times. Models are also compiled with varying options (non-quantized vs. quantized, TFLM vs. EON Compiler) to produce initial insights into RAM and flash memory usage.

Edge Impulse offers a number of tools to assist in evaluating the effectiveness of model performance. A confusion matrix can be generated from the holdout set to provide overall or per-class accuracy and F1 scores. For supported hardware, new data can be collected to perform live inference. Such evaluation options assist users in identifying trade-offs between model performance and model size and latency.

In addition to model evaluation, Edge Impulse enables post-processing evaluation and tuning using a tool known as \textit{performance calibration} \cite{situnayake2022calibration} for projects that identify events in streaming data. The tool accepts an input of user-supplied raw data or synthetically generated data along with the trained model. Using a genetic algorithm, it suggests a number of optimal post-processing configurations that trade off false acceptance rate (FAR) and false rejection rage (FRR). Suggesting optimal post-processing methods significantly reduces the engineering risk associated with a project and increases the quality of its performance.

\subsection{Compression and Optimization}

Many different optimization types can be used to improve the performance of ML and DSP algorithms when deployed to edge devices. Several types of optimization are supported by Edge Impulse, either out-of-the-box or via extensibility. The optimization areas are model compression and optimization, code optimization, and device-specific optimization.

Model compression and optimization techniques are applied either during or after training and result in models with a reduced size or computational burden when deployed to edge devices. Compression techniques available out-of-the-box in Edge Impulse include \camera{fully int-8, weight and activation} quantization \cite{benoit2017quantization} and operator fusion \cite{google2022fusion}.
\camera{Quantization-aware training is supported when converting a model to Brainchip's Neuromorphic format~\cite{businesswire2022brainchip}.}

Code optimization involves optimizations to run an algorithm or set of algorithms on a given target. This includes model-specific code generation (via EON Compiler~\cite{jongboom2020eoncompiler}), ML kernels optimized for particular processor architectures (e.g. ARM CMSIS-NN~\cite{lai2018cmsisnn}), and quantization optimized DSP algorithms. The software development kit (SDK) is designed to make use of available optimizations depending on the compiler flags that are set.

Device-specific optimizations are those that apply to specific targets due to the requirement of hardware support. Examples include the training and conversion of spiking neural networks (supported for specific targets via integration) and sparse neural networks. Additional targets and optimizations can be added using the platform’s extensibility via custom processing, learning, and deployment blocks.

Edge Impulse's EON compiler \cite{jongboom2020eoncompiler} compiles TFLM neural networks to C++ source code. The EON Compiler eliminates the need for the TFLM interpreter \camera{by generating code that directly calls the underlying kernels and enables the linker to eliminate unused instructions.} This effort reduces the RAM and ROM usage for neural network implementations, as we show 
\camera{in Section \ref{performance-eon-compiler}}.

\subsection{Conversion and Compilation}
\label{sec:conversion-and-compilation}

Edge Impulse offers several possibilities for DSP and model deployment to target embedded and edge devices, such as standalone C++ library, Arduino library, process runner for Linux, \mbox{WebAssembly} library~\cite{WebAssem60:online}, and precompiled binaries for a variety of supported boards. A deployed project includes both DSP preprocessing and trained machine learning model that have been optimized for a given architecture.

Edge Impulse provides a firmware SDK for collecting data directly on a device that will be used at inference time. This SDK can be built by a user or is available in binary format for a variety of popular microcontrollers, such as Arduino, Raspberry Pi Pico, etc. As a library, the SDK contains several public-facing functions for performing inference \cite{hymel2022inferencingsdk}. The precompiled binary presents a simple set of AT commands for usage over a serial port. 

The SDK provides a pathway to out-of-the-box operation that uses a combination of code generation, macros, and runtime checks to include more efficient algorithms and optimizations where possible, but it falls back on pure C++ where needed to run on a wide range of processor architectures. Porting to a new processor requires an allocator for the desired memory pool to service the SDK (typically a wrapper to \verb|malloc|).  \verb|printf| and timing functions help with debug and profiling, but these are not strictly necessary.

In addition to microcontrollers, the Edge Impulse inference SDK can be built and run on any Linux board utilizing x86 or ARM architectures \cite{ei2022linux}. The DSP block and trained model is downloaded from the user's project in EIM format, which is a compiled, native binary application that exposes the I/O interface for use by any number of programming languages (Python, Go, C++, Node.js, etc.).
 
\subsection{AutoML}
\label{auto-ml}

The accuracy of any deep learning system depends critically on identifying a proper choice of hyperparameters. The inherent resource constraints on embedded targets also limits the selection of hyperparameters. For example, a trade-off has to be made between allocating resources for the DSP and deep learning algorithms. For a novice user, the relationship between hyperparameters can often be difficult to grasp. Therefore, Edge Impulse provides a suite of automated machine learning (AutoML) techniques to assist non-experts in creating usable models and tune hyperparameters.

To ensure low burden on the user, Edge Impulse's EON Tuner \cite{plunkett2021eontuner} assists in the hyperparameter selection process while taking into account available RAM, ROM, and CPU clock speed of the target device. The EON Tuner helps select a number of hyperparameter configurations, including DSP preprocessing settings. It then trains the associated models to determine their accuracy. From these results, the user can select a preferred configuration (e.g. based on accuracy/F1 score or resource usage) and update the associated project to this configuration. 

\begin{figure}[btp]
    \centering
    \fbox{\includegraphics[width=.97\linewidth]{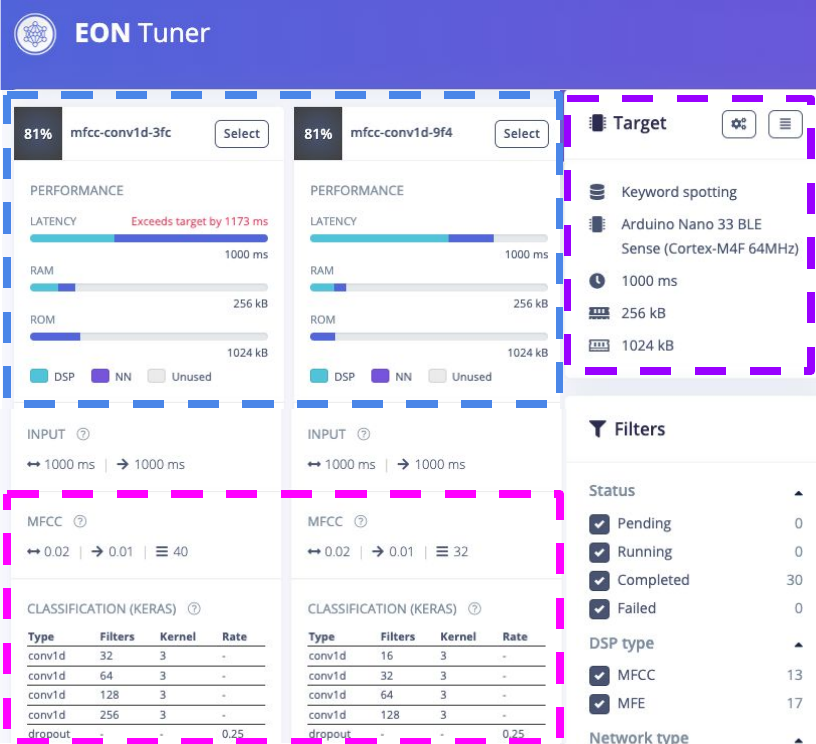}}
    \caption{\camera{Screenshot of the EON Tuner. Features are annotated with color coded dotted boxes that correspond to the challenges in Figure \ref{fig:ml-workflow}. Purple (top right): The tuner allows users to select the target hardware, which will then inform the constraints set on the search. Blue (top left): The tuner computes the configuration's accuracy and predicts the resource consumption of the DSP and NN components. Pink (bottom): The tuner searches for optimal DSP and NN combinations and displays their configuration.}}
    \label{fig:ei-eon-tuner-screenshot}

\end{figure}

Figure \ref{fig:ei-eon-tuner-screenshot} shows the user's perspective after EON Tuner has been successfully run. Note the stacked bar plots showing the estimated latency, RAM, and flash usage \camera{(Sec. \ref{sec:estimation_evaluation}) }for each combination of preprocessing (DSP) and model blocks based on the selected target (e.g. Arduino Nano 33 BLE Sense). Model details, including the specific DSP and NN configuration, are shown, which allows user to select the best combination of blocks to meet their accuracy requirements within the desired hardware constraints.

To select hyperparameter configurations, the EON Tuner combines a random search algorithm \cite{bergstra2011hyperparameter} with a heuristic to quickly estimate the performance of the configurations. Future work includes optimizing search methods using a combination of a Bayesian \cite{eggensperger2013towards} and Hyperband \cite{li2017hyperband} search algorithms. Users have the option of overriding the default search algorithm with their own search methods.

\subsection{Active Learning}

Datasets can be iteratively improved by leveraging a partially trained model to aid in labeling and data cleaning in a process called active learning~\cite{ei2022activepipeline}. Edge Impulse employs an active learning loop for the embedded sensor ecosystem where you can: (1) train a model on a small, labeled subset of your data, (2) generate semantically meaningful embeddings using an intermediate layer of the trained model, (3) visualize the embeddings (non-labeled and labeled samples) in 2D space using a dimensionality reduction algorithm (Umap~\cite{mcinnes2018umap} or t-SNE~\cite{van2008visualizing}), and (4) manually or automatically label or remove samples based on their proximity to existing class clusters.
This process can drastically speed up the labeling and data cleaning processes, which can lead to major gains in model performance.

% In addition to the tuner, Edge Impulse has support for active learning~\cite{?} to automatically identify mislabeled samples and offer suggested labels based on sample clusters \cite{ei2022activepipeline}.
% \textcolor{red}{go into more detail on activef learning (see dan's message in slack)}

\subsection{Extensibility}

% VJ: don't understand this statemnet -- Situnayake and Plunkett \cite{situnayake2023aiattheedge} define the processes and data stores that make up an MLOps deployment for edge AI, as indicated in figure \ref{fig:ml-workflow}. 

Edge Impulse supports the majority of the workflow shown in Figure~\ref{fig:ml-workflow}, with the exception of IoT device management and production monitoring. However, all Edge Impulse functionality is exposed via publicly accessible REST APIs \cite{ei2020api}, which allows users to automate the data collection, model training, and deployment processes. This API can be integrated into custom workflows and third party solutions to augment IoT device management and production monitoring, such as Microsoft Azure IoT~\cite{AzureIoT70:online}.

In addition, users are able to create their own blocks via Docker images to transform raw data taken from an existing store (e.g. AWS S3), perform feature extraction via DSP, train a custom ML model, or deploy a model. Finally, the Edge Impulse inferencing SDK library allows users to develop complete embedded ML solutions that include a variety of model compression and optimization techniques.

\camera{Futhermore, Edge Impulse can integrate with existing ML development pipelines via the Edge Impulse Python SDK~\cite{situnayake2023SDK}}. This allows users to use specific features, such as profiling or deployment (Sec. \ref{sec:estimation_evaluation} \& \ref{sec:conversion-and-compilation}), without needing to use the graphical interface.

\subsection{Scalable Infrastructure}

Edge Impulse employs AWS Elastic Kubernetes Service~\cite{ManagedK7:online} to dynamically scale compute resources based on workload requirements. All workloads are containerised, which has proven vital for efficient dependency management. Often, ML software infrastructure requires a wide-range of dependencies and versions of dependencies that are not always mutually compatible. The choice of Kubernetes over a vendor-specific tool, such as AWS Elastic Container Service~\cite{amazonecs:online}, is to enable migration of the Edge Impulse infrastructure to a different cloud provider or on-premise with a reasonable (1-6 months) amount of effort.

\section{Performance Evaluation}
\label{sec:performance}
 
ML development often narrowly focuses on the model performance in isolation~\cite{richins2021ai} due to the complexity of co-optimization (Sec. \ref{dev-deploy-challenges}).
However, the DSP stage can be a dominant factor in the overall latency and memory consumption of a TinyML application. 
Edge Impulse is designed to quantify the DSP overhead and allow users to explore the rich DSP and NN co-design space.
% Edge Impulse supports a variety of target platforms for deployment, including Arm Cortex-M, Arm Cortex-A, Tensilica Xtensa, and it also extends to edge devices like the NVIDIA Jetson series. Where possible, the Edge Impulse inference library automatically detects the target system to utilize available hardware acceleration functions. For example, this includes the CMSIS-DSP and CMSIS-NN abstraction layers for Arm Cortex-M processors and the TensorRT library for NVIDIA GPUs. 

In this section, we characterize the latency, SRAM, and flash consumption of TinyML workloads across multiple devices, optimizations, and AutoML defined configurations, thereby showing Edge Impulse's ability to address the challenges of hardware heterogeneity (Sec. \ref{hardware-het}), software fragmentation (Sec. \ref{software-frag}), and cross-stack optimization (Sec. \ref{dev-deploy-challenges}).
% we compare the performance of three different microcontroller platforms to quantify the Edge Impulse inference capabilities. Our results focus on the performance of (1) the Edge Impulse inference library and (2) demonstrating how the EON Tuner can be used to quickly narrow a search space for viable combinations of preprocessing and ML models.

\subsection{Experimental Setup}

\begin{centering}
\begin{table}[!t]
\resizebox{\linewidth}{!}{
\begin{tabular}{|c|c|c|c|c|}
    \hline
    Platform & Processor & Clock & Flash & RAM\\
    \hline
    Nano 33 BLE Sense & Arm Cortex-M4 & 64 MHz & 1 MB & 256 kB\\
    ESP-EYE (ESP32) & Tensilica LX6 & 160 MHz & 4 MB & 8 MB\\
    Ras. Pi Pico (RP2040) & Arm Cortex-M0+ & 133 MHz & 16 MB & 264 kB\\
    \hline
\end{tabular}
}
\caption{Embedded platforms used for evaluation. 
\vspace{-0.75cm}
% Each of these platforms show the extreme differences in resource constraints, as well as microprocessor instruction set architecture capabilities that are not visible in the table but our results will discuss these details.
}
\label{table:test-hardware}
\end{table}
\end{centering}

We evaluated three representative hardware designs. The details for the platforms are shown in Table \ref{table:test-hardware}. 
% The first platform is the Arduino Nano 33 BLE Sense (Nordic nRF52840) running an Arm Cortex-M4 with FPU at 64 MHz. The second platform is the Espressif ESP-EYE (ESP32) running a Tensilica LX6 at 160 MHz. Finally, the third platform is a Raspberry Pi Pico (RP2040) running an Arm Cortex-M0+ at 133 MHz. 
We chose these platforms for their differences in clock speeds, flash storage and RAM capacity.
We chose several models to evaluate the platforms to demonstrate the capabilities of each. These models were created to solve three tasks outlined in the MLPerf Tiny Benchmark \cite{banbury2021mlperf}: keyword spotting (KWS), visual wake words (VWW), and image classification. KWS is a common task in embedded devices that require wake word detection, such as “Alexa” or “OK Google.” \camera{We chose a DS-CNN model \cite{sorensen2020depthwise} that achieved at least 78\% on a test set from the Google Speech Commands dataset \cite{warden2018speechcommands}.} For the VWW task, MobileNetV1 was trained using the visual wake words dataset \cite{chowdhery2019visualwakewords}, which was derived from the Microsoft COCO dataset \cite{lin2014coco}. This dataset is a balanced set of “person” and “non-person” images used to train an image classification model. We achieved at least 72\% accuracy on a hold-out set. Finally, we trained a simple convolutional neural network (CNN) on CIFAR-10 \cite{krizhevsky2009cifar10}.

% Our benchmarks demonstrate the wall-clock time of each of the models running on each of the platforms. 100 inference tests were run on each platform for each task and the timing results were averaged. We also report the estimated RAM and flash utilization from Edge Impulse for each platform. The latency and memory utilization are compared across the for floating point (non-quantized) and INT8 quantized models with and without EON Compiler optimization. Finally, we evaluate EON Tuner on KWS to show how AutoML tools can effectively search through hyperparameters to help newcomers and experienced ML practitioners discover model architectures that meet their accuracy requirements within set timing and memory constraints.

\subsection{Cross-Hardware Inference Latency Comparison}

% Timing table
\begin{table}[t]
\centering
\resizebox{.485\textwidth}{!}{
\begin{tabular}{|c|c c|c c|c c|}
    \hline
    \multirow{2}{*}{} & \multicolumn{2}{c|}{Nano 33 BLE Sense} & \multicolumn{2}{c|}{ESP-EYE} & \multicolumn{2}{c|}{Ras. Pi Pico}\\
    \cline{2-7}
    & Float & Int8 & Float & Int8 & Float & Int8\\
    \hline
    \hline
    \multicolumn{7}{|c|}{\bf Keyword Spotting (KWS) inference times}\\
    \hline
    Preprocessing & 141.65 & 138.76 & 305.53 & 304.11 & 590.74 & 590.87\\
    \hline
    Inference & 2866.11 & 322.71 & 648.42 & 314.14 & 5700.03 & 1117.65\\
    \hline
    Total & 3007.91 & 461.62 & 954.02 & 618.35 & 6290.95 & 1708.71\\
    \hline
    \hline
    \multicolumn{7}{|c|}{\bf Visual Wake Words (VWW) inference times}\\
    \hline
    Preprocessing & - & 9.98 & 24.25 & 9.07 & - & 56.44\\
    \hline
    Inference & - & 754.74 & 2309.15 & 662.85 & - & 2205.76\\
    \hline
    Total & - & 816.56 & 2346.03 & 702.63 & - & 2286.68\\
    \hline
    \hline
    \multicolumn{7}{|c|}{\bf Image Classification (IC) inference times}\\
    \hline
    Preprocessing & 1.36 & 1.14 & 1.09 & 1.03 & 4.57 & 6.46\\
    \hline
    Inference & 1518.64 & 229.54 & 340.45 & 191.15 & 3048.05 & 554.04\\
    \hline
    Total & 1520.25 & 232.56 & 341.62 & 197.36 & 3048.05 & 561.86\\
    \hline
\end{tabular}
}
\caption{\camera{Preprocessing and inference times (in milliseconds). `-' indicates the model did not fit due to flash or RAM constraints.}
\vspace{-0.6cm}}
\label{table:timing}
\end{table}

Table \ref{table:timing} displays three sets of \textit{end-to-end} timing results using provided hardware timers. The table shows the latency of the KWS, VWW, and image classification tasks for both floating point and quantized integer (8-bit) models across the three platforms. The preprocessing and classification tasks are timed from within the Edge Impulse SDK, and the total time is taken by measuring the difference between timestamps taken around the call to run classification, which is a combination of preprocessing and inference plus some overhead not measured in either preprocessing or inference.

% \usepackage{multirow}
\begin{table*}[t]
\tiny
\resizebox{\linewidth}{!}{
\begin{tabular}{|c|c|c|ccc|ccc|ccc|}
\hline
\multirow{2}{*}{Preprocessing} &
  \multirow{2}{*}{Model} &
  \multirow{2}{*}{$\downarrow$ Acc.} &
  \multicolumn{3}{c|}{Latency (ms)} &
  \multicolumn{3}{c|}{RAM (kB)} &
  \multicolumn{3}{c|}{Flash (kB)} \\ \cline{4-12} 
 &
   &
   &
  \multicolumn{1}{c|}{DSP} &
  \multicolumn{1}{c|}{Infer.} &
  Total &
  \multicolumn{1}{c|}{DSP} &
  \multicolumn{1}{c|}{Infer.} &
  Total &
  \multicolumn{1}{c|}{DSP} &
  \multicolumn{1}{c|}{Infer.} &
  Total \\ \hline
MFE (0.02, 0.01, 40) &
  MobileNetV2 0.35 &
  85\% &
  \multicolumn{1}{c|}{332} &
  \multicolumn{1}{c|}{2420} &
  2752 &
  \multicolumn{1}{c|}{25} &
  \multicolumn{1}{c|}{468} &
  493 &
  \multicolumn{1}{c|}{-} &
  \multicolumn{1}{c|}{2242} &
  2242 \\ \hline
MFCC (0.02, 0.01, 40) &
  4x conv1d (32 to 256) &
  75\% &
  \multicolumn{1}{c|}{770} &
  \multicolumn{1}{c|}{437} &
  1207 &
  \multicolumn{1}{c|}{30} &
  \multicolumn{1}{c|}{35} &
  65 &
  \multicolumn{1}{c|}{-} &
  \multicolumn{1}{c|}{645} &
  645 \\ \hline
MFCC (0.02, 0.01, 32) &
  4x conv1d (16 to 128) &
  73\% &
  \multicolumn{1}{c|}{579} &
  \multicolumn{1}{c|}{197} &
  776 &
  \multicolumn{1}{c|}{26} &
  \multicolumn{1}{c|}{20} &
  46 &
  \multicolumn{1}{c|}{-} &
  \multicolumn{1}{c|}{221} &
  221 \\ \hline
MFE (0.02, 0.01, 32) &
  3x conv1d (32 to 128) &
  72\% &
  \multicolumn{1}{c|}{319} &
  \multicolumn{1}{c|}{174} &
  493 &
  \multicolumn{1}{c|}{21} &
  \multicolumn{1}{c|}{31} &
  52 &
  \multicolumn{1}{c|}{-} &
  \multicolumn{1}{c|}{231} &
  231 \\ \hline
MFE (0.02, 0.02, 32) &
  2x conv1d (32 to 64) &
  70\% &
  \multicolumn{1}{c|}{163} &
  \multicolumn{1}{c|}{109} &
  272 &
  \multicolumn{1}{c|}{14} &
  \multicolumn{1}{c|}{17} &
  31 &
  \multicolumn{1}{c|}{-} &
  \multicolumn{1}{c|}{125} &
  125 \\ \hline
MFCC (0.05, 0.025, 40) &
  3x conv1d (16 to 64) &
  69\% &
  \multicolumn{1}{c|}{327} &
  \multicolumn{1}{c|}{48} &
  375 &
  \multicolumn{1}{c|}{19} &
  \multicolumn{1}{c|}{10} &
  29 &
  \multicolumn{1}{c|}{-} &
  \multicolumn{1}{c|}{98} &
  98 \\ \hline
MFE (0.05, 0.025, 32) &
  2x conv1d (32 to 64) &
  69\% &
  \multicolumn{1}{c|}{161} &
  \multicolumn{1}{c|}{67} &
  228 &
  \multicolumn{1}{c|}{15} &
  \multicolumn{1}{c|}{14} &
  29 &
  \multicolumn{1}{c|}{-} &
  \multicolumn{1}{c|}{56} &
  56 \\ \hline
MFE (0.032, 0.016, 32) &
  2x conv1d (16 to 32) &
  66\% &
  \multicolumn{1}{c|}{217} &
  \multicolumn{1}{c|}{91} &
  308 &
  \multicolumn{1}{c|}{16} &
  \multicolumn{1}{c|}{19} &
  35 &
  \multicolumn{1}{c|}{-} &
  \multicolumn{1}{c|}{56} &
  56 \\ \hline
\end{tabular}
}
\caption{Preprocessing blocks and models explored with EON Tuner for the keyword spotting task. Latency, RAM, and flash estimates from EON Tuner for the keyword spotting task on the Arduino Nano 33 BLE Sense (float32 inference, using TFLM).}
\label{tab:eon-tuner-table}
\end{table*}

On some tasks, such as keyword spotting, the preprocessing time can easily equal or exceed the inference time of the unoptimized model. Therefore, optimizing the network inference via quantization, etc. will not yield the typical magnitude of latency reduction. Edge Impulse allows users to look at the end-to-end performance of a task, rather than focus only on isolated network performance.

Edge Impulse helps users experiment with preprocessing and models to quickly iterate on designs to find acceptable solutions to such problems. Table~\ref{tab:eon-tuner-table} shows how users can choose to use different preprocessing blocks, Mel-filterbank energy (MFE) or MFCCs, and sweep different model architectures with the EON Tuner for optimizing latency, accuracy, RAM, and Flash storage. There is no ideal solution; the ultimate choice is up to the user as they know their deployment constraints. Edge Impulse simply automates the possibilities and displays suggested configurations.

\subsection{Memory Optimization with EON Compiler}
\label{performance-eon-compiler}

Embedded systems are constrained by their memory and storage capacity (Section~\ref{sec:constraints}), 
% it is useful for developers to understand resource usage prior to deployment. During the feature extraction step, the Edge Impulse Studio compiles the required preprocessing code and measures the heap utilization to provide an estimate. 
% The Studio compiles the inference code and identifies the amount of heap allocated to the TFLM arena as well as the flash required.
Table \ref{table:memory-estimation} details the estimated memory usage, RAM and flash, for all three tasks. The Edge Impulse EON Compiler removes the need for the TFLM interpreter for on-device inference, thus reducing the required RAM and flash in most cases. A consistent decrease in memory utilization is seen when enabling the EON Compiler as well as quantizing to an INT8 model. Quantization can decrease the accuracy of the model due to the lower precision, but in some instances (e.g. the image classification task) it improves the accuracy due to regularization.
These optimizations do not impact the preprocessing stage.

% Memory usage
\begin{table}[!h]
\centering
\resizebox{\linewidth}{!}{
\begin{tabular}{|c|c|c|c|c|c|c|c|c|c|}
    \hline
    \multirow{2}{*}{} & \multicolumn{3}{c}{\bf Keyword Spotting} & \multicolumn{3}{|c}{\bf Visual Wake Words} & \multicolumn{3}{|c|}{\bf Image Classification}\\
    \cline{2-10}
    & RAM & Flash & Acc. & RAM & Flash & Acc. & RAM & Flash & Acc.\\
    \hline
    Preprocessing & 13.0 & - & - & 4.0 & - & - & 4.0 & - & -\\
    \hline
    FP (TFLM) & 115.8 & 148.0 & \multirow{2}{*}{78.5} & 398.4 & 904.4 & \multirow{2}{*}{81.1} & 195.8 & 107.5 & \multirow{2}{*}{70.9}\\
    \cline{1-3}
    \cline{5-6}
    \cline{8-9}
    FP (EON) & 96.8 & 106.7 & & 327.7 & 861.4 & & 162.7 & 78.7 & \\
    \hline
    Int8 (TFLM) & 38.5 & 98.1 & \multirow{2}{*}{78.5} & 124.8 & 361.2 & \multirow{2}{*}{79.9} & 51.9 & 63.1 & \multirow{2}{*}{71.1}\\
    \cline{1-3}
    \cline{5-6}
    \cline{8-9}
    Int8 (EON) & 36.4 & 65.3 & & 131.0 & 309.5 & & 44.0 & 42.1 & \\
    \hline
\end{tabular}
}
\caption{\camera{Memory estimation (all memory estimates given in kilobytes, accuracy in percentage based on the holdout set). Flash utilization estimation not provided for DSP preprocessing.}}
\label{table:memory-estimation}
\end{table}

\subsection{Design Space Exploration with EON Tuner}
% Section \ref{dev-deploy-challenges} outlines the complexity introduced by cross-stack optimization, but this process is still necessary to meet the tight constraints of TinyML systems. The EON  Tuner (Section \ref{auto-ml}) allows users to rapidly explore the complex design space to find a configuration that meets their design constraints. The Eon Tuner will search for the optimal configuration while calculating the test accuracy and predicting latency, RAM, and Flash consumption. 
Table \ref{tab:eon-tuner-table} shows a number of configurations that are searched by the EON Tuner in order to find an optimal keyword spotting model.
A user can find a model that balances the resources allocated to the DSP and NN stages in order to meet the hardware constraints of the application while maximizing accuracy. 
For example, the configurations on the 3rd and 4th lines from the bottom balance the DSP and NN differently, leading to a slightly higher accuracy and lower latency (more NN) compared to less RAM and Flash consumption (more DSP).
This process accelerates the initial exploratory phase of ML development and makes cross-stack optimizations accessible for novice ML developers.

\section{Ecosystem Enablement}

Since its launch in late 2019, Edge Impulse has seen excitement around embedded ML deployments in a variety of domains. We highlight a few exemplar use cases here.

%SlateSafety can look for early warning signs of heat exhaustion in first responders and industrial workers without an Internet connection \cite{ei2021slatesafety}. Finally, Izoelektro developed a power grid monitoring device to notify operators of potential problems \cite{ei2021izoelektro}. 

% Finally, Edge Impulse maintains active communities on forums and several social media platforms to assist users in troubleshooting technical issues.

\subsection{Education and Learning}

Edge Impulse provides both a graphical user interface through Studio as well as an extensible web-based API. Consequently, it is well suited for classroom activities, as it provides a series of parameters, plots, and visualizations within the Studio to assist newcomers in building end-to-end embedded/edge machine learning systems.
%Educators can deliver hands-on, interactive lessons without the overhead of installing and maintaining an embedded machine learning system, such as TensorFlow Lite Micro.

We saw a large interest in embedded machine learning courses upon the delivery of two separate massively open online courses (MOOCs). Between September 2020 and June 2021, over 43,000 students enrolled in three of the Tiny Machine Learning courses on EdX~\cite{TinyMach95:online}. In less than two years, over 75,000 students have benefited from these courses alone~\cite{reddi2021widening}. Additionally, over 30,000 students enrolled in the two Embedded Machine Learning courses between February 2021 and October 2022 that use Edge Impulse~\cite{Introduc11:online}. In addition, throughout 2021 and 2022, the TinyML4D Academic Network~\cite{tinymledu2022network}, which focuses on improving access to edge ML education and technology in developing countries by running a series of workshops for Africa, Asia, and Latin America regions, used Edge Impulse to teach embedded ML to professors, lecturers, and students. These 2022 workshops had 216 attendees from 48 different countries. The high attendance in the MOOCs along with the apparent growing desire for various schools to teach ML demonstrates the need for approachable tools for education.

%In addition to the EdX MOOC, Prof. Reddi developed and teaches CS249r (Tiny Machine Learning) at Harvard \cite{reddi2022cs249r}. He maintains the Harvard School of Engineering and Applied Sciences (SEAS) TinyMLedu microsite \cite{tinymledu2022welcome} that offers teaching material and course material to anyone interested in teaching embedded ML. Additionally, Edge Impulse offers open source courseware, including videos, slides, reading material, and project prompts, that combines material from the Coursera and EdX MOOCs \cite{hymel2022embeddedmlcourseware}. This material is available for anyone looking to create a course or workshop teaching edge ML. 

% Harvard SEAS partners with International Centre for Theoretical Physics (ICTP) to run the TinyML4D Academic Network \cite{tinymledu2022network}, which focuses on improving access to edge ML education and technology in developing countries. Throughout 2021 and 2022, TinyML4D ran a series of workshops for Africa, Asia, and Latin America regions to teach edge ML to professors, lecturers, and students. Edge Impulse participated in the 2022 workshops to demonstrate its toolset for developing and deploying edge ML systems. These 2022 workshops had 216 attendees from 48 different countries. The high attendance in the MOOCs along with the apparent growing desire for various schools to teach edge ML demonstrates the need for approachable tools in a classroom setting.

\subsection{Industry Use Cases, Research and Dissemination}

Edge Impulse (and indeed, edge ML) is being used by a number of companies to solve unique problems. For example, Oura Ring, which uses ML models to identify sleep patterns \camera{(Sec. \ref{sec:oura-ring})}, relies on Edge Impulse. Edge Impulse is also being adopted by academic researchers as a tool for enabling scientific research and rapid problem-solving using ML. Since Edge Impulse simplifies the machine learning training and deployment process, it allows researchers to focus on solving the problem within their domain expertise rather than setting up an end-to-end system using another framework and dealing with low-level details. For example, recent work shows how to easily develop an ``intelligent chemical sniffer, capable of detecting hazardous volatile organic compounds''~\cite{shamim2022vocs}. Another recent work identifies cancerous growths on oral tongue lesions using an embedded device~\cite{shamim2022tongue}. Other areas of research include creating low-power solutions to classify mosquito wingbeat sounds in order to identify mosquito species~\cite{altayeb2022mosquito}. There are several other such examples that are enabling non-ML scientists to easily adopt Edge Impulse to realize edge ML deployments.

\subsection{Open Source, Community Development}

Developing the open source embedded ML community is a must for the success of the TinyML industry. To this end, Edge Impulse maintains many open source repositories on GitHub~\cite{ei2022github}, including repositories related to machine learning, device firmware, and sample use-cases. As a result, there is a community of users who contribute code, issues, and bug-fixes beyond the Edge Impulse team.

This emphasis on community and open-source principles carry 
% through to other aspects of the platform as well, 
beyond the GitHub repositories. For example, one feature intended to share knowledge is the concept of \textit{public projects}, where developers can make their work inside the Edge Impulse Studio public for other developers to review and clone, thus helping share applied techniques and best practices in Edge Impulse projects. When a project is set to public, it is also aggregated and searchable via the Projects page~\cite{ei2022projects} on the Edge Impulse website. This searchable index allows developers to sort, filter, and search for relevant examples and public work. At the time of writing, there are 3,219 public Edge Impulse projects available.

In addition to public projects, preprocessing and model inference libraries (along with the trained model) may be downloaded from Edge Impulse and shared via open source licenses (Apache 2.0, BSD 3-Clause, MIT). The ability to share projects and code allows researchers and developers to disseminate their trained ML systems for reproducibility.

Collaboration can also occur inside of an Edge Impulse project. Through the use of \textit{Organizations}, more than one developer can access or participate in the data upload, analysis, learning block, model creation and testing, or other aspects of the platform or project. Having multiple users can speed up project delivery of course, but the ability to spread knowledge and share best practices can positively impact users who might be starting out with edge ML. 

\newcommand{\greencheckmark}{{\color{OliveGreen}\cmark}}
\newcommand{\redxmark}{{\color{BrickRed}\xmark}}

\begin{table}[t!]
\centering
\resizebox{\linewidth}{!}{
\begin{tabular}{|c|c|c|c|c|c|}
 \hline
 & 
 
 \shortstack{Data \\ Collection \\ \& Analysis} & 
 \shortstack{DSP \& \\ Model \\ Design} & 
 \shortstack{Embedded \\ Deployment \newline } &
 \shortstack{AutoML \\ \& Active \\ Learning} & 
 \shortstack{ IoT \\ Management \\ \& Monitoring}
 \\ [0.5ex] 
 \hline
 
\hline\hline
 \textbf{Edge Impulse} & \greencheckmark & \greencheckmark & \greencheckmark & \greencheckmark & 
  \color{BurntOrange}$\sim$ \\
\hline\hline
 Amazon SageMaker  & \greencheckmark & \color{BurntOrange}$\sim$ & 
 \color{BurntOrange}$\sim$ & 
 \greencheckmark &
 \color{BurntOrange}$\sim$ \\

\hline
 Google VertexAI & \greencheckmark & \color{BurntOrange}$\sim$ & 
 \redxmark & 
 \greencheckmark &
 \color{BurntOrange}$\sim$ \\

\hline
 Azure ML \& IoT & \greencheckmark & \color{BurntOrange}$\sim$ & \color{BurntOrange}$\sim$ & \greencheckmark & \greencheckmark \\

\hline
 Neuton AI & \redxmark & \color{BurntOrange}$\sim$ & \ \greencheckmark & \color{BurntOrange}$\sim$ &
 \redxmark \\

\hline
 Latent AI & \redxmark & \greencheckmark & \ \greencheckmark & \redxmark &
 \redxmark \\

\hline
 NanoEdge & \color{BurntOrange}$\sim$ & \greencheckmark & \ \greencheckmark & \color{BurntOrange}$\sim$ &
 \redxmark \\

\hline
 Imagimob & \greencheckmark & \greencheckmark & \greencheckmark & \color{BurntOrange}$\sim$ &
 \redxmark \\
 
 \hline

\end{tabular}
}
% \vspace{12pt}
\caption{\camera{Comparison of supported features of MLOps platforms. \greencheckmark:Fully Supported, {\color{BurntOrange}$\sim$}:Partially Supported, \redxmark:Not Supported. }}
\vspace{-0.5cm}
\label{table:realted-platforms}
\end{table}

\section{Related Work}
\label{sec:related}

MLOps platforms like Amazon SageMaker~\cite{rauschmayr2021amazon}
% ,joshi2020amazon}
and Google VertexAI~\cite{VertexAI23} exist, but they cater to cloud-scale environments that are vastly different from the TinyML ecosystem. Microsoft's Azure IoT~\cite{klein2017iot}
% ,AzureIoT70:online}
is a platform specifically designed to connect, analyze, and automate data analytics from the edge to the cloud. Edge Impulse uniquely encompasses a large slice of the embedded ecosystem. The platform operates as a dataset analysis and preprocessing tool, a training framework, an optimizer, and an inference engine. Additionally, it readily integrates with other cloud services.

TinyMLOps frameworks focus on the ML training and compression stages of the ML pipeline~\cite{leroux2022tinymlops}, while Edge Impulse spans data collection to deployment. Neuton AI~\cite{NeutonAI32:online} allows users to train a TinyML model from CSV data, but it does not include features for data collection and limits the customization of the preprocessing and model architecture. Latent AI~\cite{LatentAI69:online} provides a platform to compress and compile a model for efficient deployment but does not support data collection and training workflows. NanoEdge AI Studio~\cite{NanoEdgeAIStudio} is a TinyML development tool primarily focusing on time series data exclusively for STM devices. Imagimob~\cite{imagimob2022imagimob} offers end-to-end TinyML development with various preprocessing options, AutoML for model selection, and model training, evaluation, and deployment to optimized C code. However, it does not offer a web-based API that allows for full MLOps automation. \camera{Table \ref{table:realted-platforms} illustrates the level of support the related MLOps platforms have for each set of relevant features.} 

TFLM~\cite{david2021tensorflow} is the standard inference engine for TinyML use cases because it is flexible, open-source, and associated with the TensorFlow ecosystem. Edge Impulse's EON Compiler uses less memory and storage compared to TFLM (Section \ref{performance-eon-compiler}). Arm’s uTensor~\cite{uTensoru41:online} and Microsoft’s ELL~\cite{TheEmbed23:online} are both TinyML inference engines that cross-compile to specific target embedded hardware, but both projects are no longer supported. TinyEngine~\cite{mcunet} and uTVM~\cite{microTVM31:online} compile the network graph and perform inter-layer optimizations. TinyEngine is a research project and therefore it is not designed for production scale and uTVM currently supports very few boards. Vendor-specific inference engines, such as STM32Cube.AI~\cite{XCUBEAIA36:online}, provides optimized inference but only support vendor-specific hardware, thus limiting application portability. All of these engines, however, can be adopted and integrated into Edge Impulse.

\section{\camera{Industry Case Studies}}

This section focuses on real-world insights from two industry case studies: sleep 
tracking with the Oura Ring and detecting heat exhaustion with SlateSafety Band V2.

\subsection{Oura Ring}
\label{sec:oura-ring}
The Oura Ring is designed to track the user's sleep patterns using an ML model that predicts the stages of sleep based on measured physiological signals. In order to improve their model, Oura conducted a large-scale sleep study ($\sim$100 participants) and collected a large dataset for model training.

\subsubsection{Challenge}
Incomplete, noisy, and inconsistent data is unavoidable when collecting real-world datasets and, given the scale of the study, it would be an arduous process to aggregate, scrub, and analyze the data before using it to train the next generation sleep tracking model.
Many current platforms provide little assistance for the critical stages before training, and stand-alone data analysis and visualization tools require significant additional effort to set up ad hoc ML pipelines.

\subsubsection{Edge Impulse Solution}
Edge Impulse is able to ingest and align data from multiple sources and sensors by comparing sensor signatures, which simplifies and speeds up a typically manual and error-prone process.
EI integrates analysis tools that enable domain experts to make design decisions on which sensors and features are meaningful for sleep prediction.
Through this process, Oura created a model focusing only on heart rate, motion, and body temperature and achieves a best-in-class 79\% correlation accuracy when compared to polysomnography and human scorers, which use expensive measurements, such as brain waves~\cite{dezambotti2019sleep}.

\subsubsection{Focus Areas for Future ML Systems Research}
Current ML Systems researchers too often ignore data-centric techniques and instead focus on stages, like model design, with easier-to-quantify metrics such as accuracy and latency~\cite{mazumder2022dataperf}.
Areas for high-impact systems research include:
(1) Aggregating and aligning data from unstructured sources and multiple sensors.
(2) Visualizing statistically meaningful correlations  of data from multiple sensors without hallucinations.
(3) Leveraging non-technical domain experts for data cleaning and selection.

\subsection{SlateSafety}
The SlateSafety BAND V2 is a wearable device that monitors physiological signals of first responders and industrial workers.
Due to the lack of reliable wireless connection in the deployment environment, SlateSafety required on-device inference to predict heat exhaustion in users.

\subsubsection{Challenge}
SlateSafety aimed to leverage existing hardware that was already in the field instead of going through the expensive process of developing an entirely new, ML-capable platform. Therefore, the resulting model had to run in real-time on an existing microcontroller with limited memory capacity.

\subsubsection{Edge Impulse Solutions}
Edge Impulse's EON Tuner and Compiler were able to automatically design a custom model and deploy it efficiently to the existing microcontroller via an automatic over-the-air update.
This means that SlateSafety could deploy a new feature to its users seamlessly and without a long development cycle around new hardware.

\subsubsection{Focus Areas for Future ML Systems Research}
Much of existing ML research target state-of-the-art hardware that has been designed with ML deployment in mind. However, in practice, ML-based features are deployed to existing systems.
In order to enable broader adoption of AI and prevent existing hardware from being replaced and thrown out~\cite{prakash2023tinyml}, ML researchers need to develop optimization techniques that can be backward compatible with past generations of hardware.

\section{Conclusion}
\label{sec:conclusion}

Edge Impulse is a framework focused on building machine learning systems for resource-constrained devices. 
% The graphical front-end provides an accessible on-ramp for both embedded engineers and machine learning practitioners to develop and deploy ML projects to such devices. Additionally, the web-based API allows for MLOps systems to be constructed to automate the process of data collection, model training, and model deployment.
\camera{The framework is built around the principles of accessibility, data-centric co-optimization, and cross-stack collaboration.
As a result, Edge Impulse has put AI in the hands of the individuals it impacts by reducing the expertise and computing resources required to participate. Edge Impulse has already been used in various industrial, research, and educational applications, and the lessons learned from these deployments can focus future systems research on high-impact problems.}

% We believe there are myriad possibilities to extend and grow Edge Impulse. BrainChip announced on-chip training with its Akida line of neuromorphic processors \cite{businesswire2022brainchip}. A partnership between Edge Impulse and Brain Chip \cite{brainchip2022partnership} means that development on neuromorphic systems could be made more accessible to engineers, programmers, and students. Predictive maintenance is an important area of research and development within Industrial IoT (IIoT), and anomaly detection often plays a large role in predictive maintenance systems \cite{ran2019predictivemaintenance}. As such, Edge Impulse could develop extensible ML frameworks beyond TensorFlow that allow users to develop and deploy such systems. Finally, with a focus on edge devices, Edge Impulse is poised to offer complete MLOps solutions (whether alone or through partnerships) to automatically collect data, train a new model, and update device firmware to allow such devices to adapt to their environments.

% Future work for Edge Impulse will include improving device compatibility, offering additional ML models beyond neural networks, and building robust MLOps capabilities. We see a need in industry and academia to offer accessible edge machine learning for newcomers and experienced developers with robust support for heterogeneous platforms and architectures.

%\input{tex/appendix}

% \bibliography{refs}
% \bibliographystyle{abbrv}

\bibliographystyle{mlsys2023}
\bibliography{refs}

%%%%%%%%%%%%%%%%%%%%%%%%%%%%%%%%%%%%%%%%%%%%%%%%%%%%%%%%%%%%%%%%%%%%%%%%%%%%%%%
%%%%%%%%%%%%%%%%%%%%%%%%%%%%%%%%%%%%%%%%%%%%%%%%%%%%%%%%%%%%%%%%%%%%%%%%%%%%%%%
% SUPPLEMENTAL CONTENT AS APPENDIX AFTER REFERENCES
%%%%%%%%%%%%%%%%%%%%%%%%%%%%%%%%%%%%%%%%%%%%%%%%%%%%%%%%%%%%%%%%%%%%%%%%%%%%%%%
%%%%%%%%%%%%%%%%%%%%%%%%%%%%%%%%%%%%%%%%%%%%%%%%%%%%%%%%%%%%%%%%%%%%%%%%%%%%%%%
% \appendix
% \clearpage
% \input{tex/appendix}

%%%%%%%%%%%%%%%%%%%%%%%%%%%%%%%%%%%%%%%%%%%%%%%%%%%%%%%%%%%%%%%%%%%%%%%%%%%%%%%
%%%%%%%%%%%%%%%%%%%%%%%%%%%%%%%%%%%%%%%%%%%%%%%%%%%%%%%%%%%%%%%%%%%%%%%%%%%%%%%

\end{document}